\documentclass[journal,10pt]{IEEEtran}
\usepackage[T1]{fontenc}
\usepackage{amsmath,amssymb,amsthm}
\usepackage{graphicx}
\usepackage{capt-of}
\usepackage{cite}
\usepackage[hidelinks]{hyperref}
\usepackage{eso-pic}
\hypersetup{pdftitle={Certified Outage Bounds for RIS-Assisted NOMA with Source-Side Thermal Energy Harvesting},
 pdfauthor={Victor Rabiet},
 pdfkeywords={Reconfigurable intelligent surface, NOMA, outage probability, Poisson interference, interval arithmetic}}

\newcommand{\E}{\mathbb E}
\newcommand{\Prb}{\mathbb P}
\newcommand{\calL}{\mathcal L}
\DeclareMathOperator{\acosh}{acosh}
\title{Certified Outage Bounds for RIS-Assisted NOMA\break with Source-Side Thermal Energy Harvesting}
\author{Victor~Rabiet\thanks{V. Rabiet is with the \'Ecole normale
sup\'erieure, PSL University, 75005 Paris, France, and with LTCI,
T\'el\'ecom Paris, Institut Polytechnique de Paris, 91120 Palaiseau,
France (e-mail: victor.rabiet@telecom-paris.fr).}}
\begin{document}
\AddToShipoutPictureFG*{\AtPageUpperLeft{\put(\LenToUnit{0.5\paperwidth},\LenToUnit{-0.28in}){\makebox[0pt][c]{\parbox{0.92\paperwidth}{\centering\scriptsize\itshape This work has been submitted to the IEEE for possible publication. Copyright may be transferred without notice, after which this version may no longer be accessible.}}}}}
\maketitle

\begin{abstract}
Gaussian useful-gain and mean-interference approximations can cause
opposite power-sizing errors in reconfigurable intelligent surface (RIS)-assisted non-orthogonal multiple access (NOMA). We extend a source-side thermal energy harvesting model with Poisson interference and certify two-sided outage bounds. In a tested environment, the combined required signal-to-noise ratio error has opposite signs at 20 and 24 elements per user.
Under another power convention, mean or moment-matched Gamma-distributed interference
can predict false feasibility with exact useful gain. In the Gamma case, the Gaussian gain
approximation can mask this false feasibility. Propagated
Chernoff bounds give sufficient designs. Results concern the specified
model and tested regimes.
\end{abstract}


\section{Introduction}
\IEEEPARstart{S}{mall} outage targets make the direction of an approximation
error a design issue. A conservative useful-channel approximation may
hide an optimistic interference approximation, so apparent agreement
need not justify a reliable design. We study this interaction in the reconfigurable intelligent surface (RIS) and non-orthogonal multiple access (NOMA) model of Boujemaa \emph{et al.}\ \cite{boujemaa}, whose source is thermally powered. We use their squared-Gaussian useful-gain approximation \cite[Eq.~(22)]{boujemaa}, denoted \eqref{eq:gaussian}, as a comparator and
extend the model with independent Poisson interference. The metric is
outage at a fixed decoding threshold. Taken separately, \eqref{eq:gaussian} and the
mean-interference rule are expected to fail at rare-event targets. We
study how their errors combine.
Ultra-reliable low-latency communication (URLLC) motivates the probability
targets, but we give no packet-level latency guarantee.

Prior work gives exact RIS gain distributions and finite bounds.
Atapattu \emph{et al.}\ \cite{atapattu} give two-sided bounds and Gamma
approximations, and Gamma-fitted composite fading underlies
stochastic-geometry analyses of RIS-aided NOMA networks \cite{xie}. Generalized-fading expressions and exact Rayleigh-product
sums appear in \cite{trigui,luna}. Wang \emph{et al.}\ \cite{wang} give
the Laplace transform, a Chernoff bound and saddlepoint approximations,
and identify limitations of Gaussian tails. Absolute-error analyses
\cite{bulla,anade} and series with truncation control \cite{garcia}
also guide the choice of the approximation. Poisson interference is standard
shot-noise modeling \cite{baccelli}, but its transform alone does not fix
the numerical accuracy of a rare-outage calculation. Explicit Gaussian
bounds for Poisson shot noise \cite{inaltekin,khabou} carry their own
guarantees, and dominant-interferer-plus-mean closures \cite{qin} are
validated numerically. We evaluate the full Poisson point process (PPP) interference distribution with a numerical error budget.

Relative to these works, our two-sided certificate covers thermal averaging, full PPP interference and rounding, gives Chernoff margins in signal-to-noise ratio (SNR) and proves all-SNR infeasibility missed by the largest-interferer bound. We separate useful-gain and
interference errors. When interferer powers stay fixed in noise units, mean interference overestimates outage at sufficiently high SNR.

\section{System Model}
A thermally powered source serves two NOMA users, $i=1,2$, through an RIS, with no direct path. Each user is assigned a disjoint sub-surface of $N$ elements ($2N$ total), ideally phase-aligned with perfect channel state information, at fixed per-element gain. Their outages are evaluated separately, without jointly sizing a shared surface. Let
\begin{equation}
 A_N=\sum_{k=1}^N\sqrt{E_kF_k},\qquad
 W=\frac{A_N^2(10+Z)^2}{101},
 \label{eq:gain}
\end{equation}
where the Rayleigh-hop power gains $E_k,F_k$ are independent $\mathrm{Exp}(1)$ variables and
$Z\sim\mathcal N(0,1)$ is independent of them. The cumulative distribution functions (CDFs)
are $F_N(a)=\Prb(A_N\le a)$ and $F_W(w)=\Prb(W\le w)$.
With harvested transmit energy $\zeta(10+Z)^2$, deterministic $\zeta>0$ and noise level $N_0$, the average-energy SNR is $\rho=101\zeta/N_0>0$, plotted as $10\log_{10}\rho$. The thermal factor $(10+Z)^2/101$ has unit mean.

Power fractions $p_1=0.3$, $p_2=0.7$ and decoding threshold $\theta=1$
give $c_1=\max\{\theta/p_1,\theta/(p_2-\theta p_1)\}=10/3$
with perfect successive interference cancellation, and
$c_2=\theta/(p_2-\theta p_1)=5/2$.
Useful path gains $\ell_1=1/8$, $\ell_2=1/64$ give
$d_i=c_i/\ell_i$, hence $d_1=80/3$, $d_2=160$.
These gains remain fixed across interference path-loss exponents.
The thermal mean $10$ and $p_1$ follow \cite[Sect.~4]{boujemaa}; the unit thermal standard deviation and the useful path-loss exponent $3$ follow the code of \cite[App.~B]{boujemaa}; the path gains, $\theta$ and
the interference levels below are choices of this benchmark.

Let $\Pi$ be a planar PPP of intensity $\lambda>0$
outside the guard disk of radius $R_0=1$ centered on user $i$
(distances are normalized by $R_0$), and let
$J=\sum_{x\in\Pi}H_x|x|^{-\eta}$, $\eta>2$.
The $H_x$ are independent unit-mean exponential \emph{power} marks,
and the marked PPP is independent of the useful fading and thermal factor.
Interference reaches user $i$ directly, not through the RIS. Under the constant interference-to-noise ratio convention (C-INR), interferer powers stay fixed in noise units. The transmit-power convention (C-TX) assigns interferers the source's mean transmit power, and their exponential marks remain independent of the useful-link thermal factor. With noise and the interference path gain at distance $R_0$ normalized to
one, so that in C-INR a unit-mark interferer at the guard distance is
received at the noise level, the outage probabilities of user $i$ are
\begin{align}
 P_{i,\mathrm{INR}}(\rho)&=\E F_W\bigl(d_i(1+J)/\rho\bigr),\nonumber\\
 P_{i,\mathrm{TX}}(\rho)&=\E F_W\bigl(d_i(J+\rho^{-1})\bigr).
 \label{eq:outages}
\end{align}
The two conventions describe different experiments. C-INR obeys $P_{2,\mathrm{INR}}(6\rho)=P_{1,\mathrm{INR}}(\rho)$; C-TX does so only for $J=0$.

Writing $m_N=N\pi/4$ and $s_N^2=N(1-\pi^2/16)$, the squared-Gaussian comparator is
\begin{equation}
 H_N(a)=\Phi\!\left(\frac{a-m_N}{s_N}\right)
       -\Phi\!\left(\frac{-a-m_N}{s_N}\right),
 \label{eq:gaussian}
\end{equation}
where $\Phi$ is the standard normal CDF. Both branches are retained,
so $H_N(0)=0$. Replacing $F_N$ by $H_N$ changes only the useful gain,
and replacing $J$ by $\E J$ only the interference.

\section{Two-Sided Outage Certification}
\subsection{Gain Inversion and Thermal Integration}
For $V=\sqrt{E_1F_1}$, $f_V(a)=4aK_0(2a)$, with $K_0$ the modified
Bessel function. Its known transform \cite{wang} is
\begin{equation}
 \calL(s)=\frac{q\acosh(q)/\sqrt{q^2-1}-1}{q^2-1},\quad q=s/2,
 \label{eq:laplace}
\end{equation}
continued from $s>2$ to $\Re s>-2$, with $\calL(2)=1/3$.
We set $B_A=N+16\lceil\sqrt N\rceil+40$, $M_A=2^{19}$,
$c_AB_A=20$ and $a_0=(e^{20}-1)^{-1}$.
As in the Fourier-series method of Abate and Whitt \cite{abate}, the
periodic inverse of $\calL(s)^N/s$ satisfies
\begin{equation}
 S_A(a)=\sum_{l\ge0}e^{-20l}F_N(a+lB_A),\quad 0<a<B_A.
 \label{eq:periodic}
\end{equation}
At $a_j=jB_A/M_A$, let $S_{A,M}(a_j)$ be the normalized inverse
discrete Fourier transform (DFT) of $\calL(z_k)^N/z_k$, $z_k=c_A+2\pi\mathrm i k/B_A$,
multiplied by $M_Ae^{c_Aa_j}/B_A$. Negative frequencies are conjugated,
and the Nyquist coefficient is its real part. Then, at these nodes,
\begin{align}
 F_N(a)&\in S_{A,M}(a)-a_0+[-R_A(a),R_A(a)]
                  +[0,\tau_Aa_0],\nonumber\\
 R_A(a)&=\frac{2e^{c_Aa}}{\pi N}
       \left(\frac{(2\pi+c_A)B_A}{\pi M_A}\right)^N,
 \quad \tau_A=e^{-B_A}\calL(-1)^N .
 \label{eq:gain-certificate}
\end{align}
The last interval is one-sided: after subtracting $a_0$, periodization
leaves discounted upper tails, bounded by exponential Markov.
Integration by parts and $\int|f_V'|\le2\pi$ give
$|\calL(c_A+\mathrm it)|\le(2\pi+c_A)/|t|$.
For even $M_A\ge N$, the omitted frequency sum, including the missing
half Nyquist pair, gives $R_A$. We use $F_N(0)=0$ and
$F_N(a)\in[1-\tau_A,1]$ for $a\ge B_A$.
Between nodes, we use the left lower and right upper CDF bounds.
The discretization bound $a_0$ is from \cite{abate}. We add the
explicit truncation bound $R_A$ and outward rounding.

For $w>0$, we partition $[-12,12]$ into dyadic cells $I_j$, with
$-10$ and $0$ as boundaries. Let $m_j^\pm$ enclose their normal
probabilities by density rectangles, and let $a_j^\pm$ enclose
$\sqrt{101w}/|10+I_j|$. Monotone CDF envelopes give
\begin{equation}
 \sum_jm_j^-\underline F_N(a_j^-)
 \le F_W(w)\le
 \sum_jm_j^+\overline F_N(a_j^+)+\frac{2\phi(12)}{12},
 \label{eq:thermal}
\end{equation}
where $\phi$ is the normal density. A cell touching $-10$ has upper
argument $+\infty$, with CDF bound one, and at $w=0$ the stored enclosure contains the exact value
$F_W(0)=0$.
Stored enclosures of $F_W$ (thermal banks) are extended between and beyond their arguments by monotone endpoint bounds.

\subsection{Poisson Interference Enclosure}
For the specified PPP, its exponential formula \cite[Prop.~2.6]{baccelli}
gives
\begin{equation}
 {\calL}_J(s)=\exp\!\left[-2\pi\lambda\int_1^\infty
              \frac{sr}{r^\eta+s}\,dr\right],\quad \Re s>0 .
 \label{eq:ppp-laplace}
\end{equation}
For $\eta=4$ the exponent is $-\pi\lambda\sqrt{s}\arctan\sqrt{s}$,
with principal branches. Let $\alpha=2/\eta$ and
$\beta=\pi^2\lambda\alpha/[2\sin(\pi\alpha/2)]$.
Subtracting the unit-disk exponent from the full-plane exponent yields
$|{\calL}_J(c+it)|\le\exp(\pi\lambda-\beta|t|^\alpha)$.

We set $B=128$, $cB=20$, $M=2^{20}$, $K=M/2$, and
$\xi=\beta(2\pi K/B)^\alpha$. The absolute Fourier remainder,
including the missing half Nyquist pair, is bounded by $e^{cx}R$,
\begin{equation}
 R=\frac{e^{\pi\lambda-\xi}}{\pi}
        \left(\frac1{2K}+\frac1{\alpha\xi}\right).
 \label{eq:ppp-frequency}
\end{equation}
The frequency tail follows from the exponential-integral bound
$E_1(\xi)/\alpha\le e^{-\xi}/(\alpha\xi)$.
Let $S_M(x)$ be the real normalized inverse DFT of
$\calL_J(c+2\pi ik/B)/(c+2\pi ik/B)$, multiplied by
$Me^{cx}/B$, with the same conjugation and Nyquist conventions.
For $F_J(x)=\Prb(J\le x)$, periodization gives $S(x)=\sum_{l\ge0}e^{-20l}F_J(x+lB)$.
Since $\log\E e^{J/2}\le\mu_J=\E J=2\pi\lambda/(\eta-2)$,
exponential Markov gives
$\Prb(J>B)\le\tau=e^{-B/2+\mu_J}$. Thus, at the DFT nodes
$x=jB/M$, $0<x<B$,
\begin{equation}
 \begin{split}
 S_M-a_0-e^{cx}R &\le F_J(x)\\
                 &\le S_M-a_0+e^{cx}R+\tau a_0 .
 \end{split}
 \label{eq:ppp-certificate}
\end{equation}
We use $F_J(0)=0$ and $F_J(B)\in[1-\tau,1]$.

We transfer the DFT bounds monotonically to
$x_j=10^{j/2560}-1<B$, append $B$, and tighten to
$L_j\le F_J(x_j)\le U_j$.
CDF steps $U_{j+1}$ and $L_j$ on $[x_j,x_{j+1})$ define
$X^-\le_{\rm st}J\le_{\rm st}X^+$, with residual masses placed
at $B$ and $+\infty$, respectively. Outward abscissa bounds preserve
this order. Therefore every bounded increasing
test satisfies
\begin{equation}
 \E g(X^-)\le\E g(J)\le\E g(X^+).
 \label{eq:stochastic}
\end{equation}
Here $g(x)=F_W(d_i(1+x)/\rho)$ or $F_W(d_i(x+\rho^{-1}))$,
and $g(+\infty)=1$. Certified gain CDFs and Gaussian-density rectangles
bound the thermal integral by \eqref{eq:thermal}.
This argument requires only monotonicity, without interpolation: it also evaluates the C-TX floor directly at
$\rho^{-1}=0$, although $F_W$ is not differentiable at zero.

All constants, special functions, DFTs, masses and sums use 160-bit real/complex midpoint-radius enclosures (Arb/Acb) \cite{arb}, through python-flint~0.9.0.
Node positions and grid choices may be made in floating point; every
certifying evaluation and stored binary64 endpoint is enclosed or rounded
outward, with conversions checked in Arb. The two Rayleigh total-error signs and the $N=20$ C-TX floor
decision in Section~V were reproduced using a separate
multiple-precision floating-point interval (MPFI) implementation \cite{mpfi}. Envelopes are intersected with $[0,1]$
and residual masses are retained. The resulting interval $[P_-,P_+]$
contains no interference-substitution error.

\section{Decision Rules and Comparators}
\subsection{Certified Decisions and C-TX Floor}
At target $p^*$, $P_+\le p^*$ certifies feasibility,
$P_->p^*$ certifies infeasibility at that SNR, and otherwise the evaluation is undecided. The last excluded and first feasible SNRs enclose the required SNR (threshold), and undecided points between them are retained.
For C-TX, dominated convergence gives
\begin{equation}
 P_{i,\mathrm{TX}}(\rho)\downarrow P_{i,\mathrm{floor}}
 :=\E F_W(d_iJ),\qquad \rho\to\infty.
 \label{eq:floor}
\end{equation}
A floor lower bound above $p^*$ proves infeasibility at every SNR, which a finite SNR search cannot. An upper outage bound above
target cannot alone prove infeasibility.

\begin{figure*}[t]
\centering
\includegraphics[width=\textwidth]{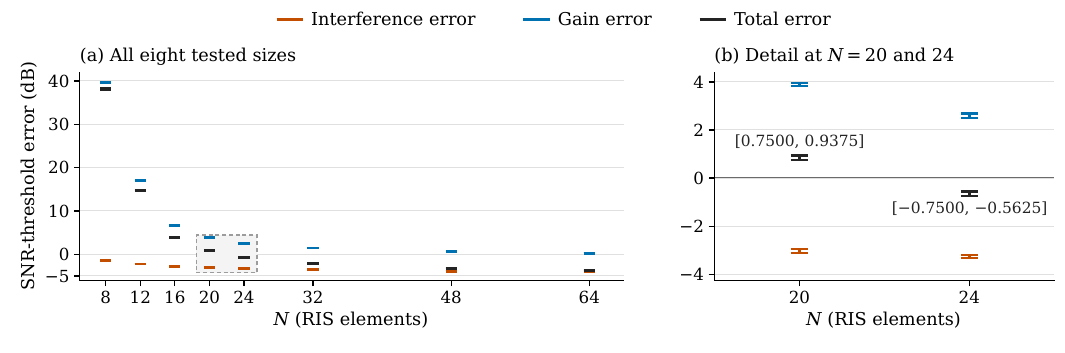}
\caption{C-INR, $\lambda=0.1$, $\eta=4$, user 1, target $10^{-5}$:
certified SNR-threshold errors \eqref{eq:errors}, drawn as interval
endpoints at the eight tested sizes; the total is enclosed directly.
Panel (b) resolves the signs at $N=20$ and $24$. All thresholds shown are
finite; no result between tested sizes is implied.}
\label{fig:errors}
\end{figure*}

\subsection{Propagated Chernoff Bound}
We use the existing Rayleigh-product bound \cite{wang}
\begin{equation}
 C_N(a)=\inf_{s\ge0}e^{sa}\calL(s)^N\ge F_N(a).
 \label{eq:chernoff}
\end{equation}
We enclose each interior minimizer between points where the objective derivative is certified negative on the left and positive on the right. Near $s=2$, we evaluate $\calL$ using series with bounded remainders. Substituting $C_N$ for $F_N$ in the same thermal
and PPP integrations encloses the propagated Chernoff comparator.
An upper endpoint $\le p^*$ certifies feasibility; a lower endpoint $>p^*$ excludes candidates under the Chernoff rule only.

Without PPP, 12 user-1 comparisons ($N=8,16,32,64$; targets
$10^{-4},10^{-5},10^{-6}$) give Chernoff grid penalties of $0.50$--$1.75$ dB,
whereas the excess of \eqref{eq:gaussian} exceeds $28$ dB for Rayleigh $N=8$ at $10^{-5}$
(Table~\ref{tab:baseline}).

The Nakagami rows use $E_k,F_k\sim\mathrm{Gamma}(m,\mathrm{rate}\ m)$,
moment-matched \eqref{eq:gaussian} and $\calL^{(2m-2)}(s/m)/\Gamma(m)^2$,
valid for the integer values $m=2,3$ considered here.

\par\smallskip\noindent\begin{minipage}{\columnwidth}
\captionof{table}{Without PPP: user 1, target $10^{-5}$.
Required SNR in dB on the $0.25$-dB grid.}
\label{tab:baseline}
\centering\footnotesize
\setlength{\tabcolsep}{4.1pt}
\renewcommand{\arraystretch}{1.10}
\begin{tabular}{lrrrr}
\hline
Fading & $N$ & Exact & \eqref{eq:gaussian} & Chernoff\\
\hline
Rayleigh & 8 & $12.00$ & $>40$ & $13.75$\\
Rayleigh & 64 & $-13.50$ & $-13.25$ & $-13.00$\\
Nakagami $m=2$ & 8 & $6.50$ & $11.75$ & --\\
Nakagami $m=3$ & 8 & $4.75$ & $6.50$ & --\\
\hline
\end{tabular}
\par\smallskip
\begin{minipage}{\columnwidth}\footnotesize
Finite entries are certified grid minima, the preceding point being
excluded; $>40$ is censored at the search limit; -- means not evaluated.
Chernoff entries concern its sufficient rule.
\end{minipage}
\end{minipage}\par\smallskip

For C-INR, $\lambda=0.1$, $\eta=4$, user~1 and $p^*=10^{-5}$,
the exact thresholds lie in $[3.1250,3.2500]$ dB at $N=20$ and
$[1.1875,1.2500]$ dB at $N=24$.
The propagated Chernoff-rule brackets are $[4.3125,4.4375]$ and
$[2.3125,2.3750]$ dB, respectively. Their upper endpoints certify sufficient SNRs; the Chernoff
thresholds exceed the exact ones by $[1.0625,1.3125]$ and
$[1.0625,1.1875]$ dB.

\subsection{Benchmark Protocol and Error Definitions}
The PPP benchmark uses eight sizes ($N=8$, $12$, $16$, $20$, $24$,
$32$, $48$, $64$), $\lambda\in\{0.1,1\}$, $\eta\in\{3,4\}$,
both users, both power conventions and targets
$10^{-3},10^{-4},10^{-5}$: 384 cases.
Each case compares five models: exact gain with PPP, mean or Gamma interference, and \eqref{eq:gaussian} with mean or Gamma interference.
Gamma interference uses shape $\mu_J^2/\sigma_J^2$ and scale $\sigma_J^2/\mu_J$, with $\mu_J=\E J=2\pi\lambda/(\eta-2)$ and $\sigma_J^2=\operatorname{Var}(J)=4\pi\lambda/(2\eta-2)$.
Both Gamma comparators are evaluated rigorously, but their enclosures
alone do not bound the PPP outage.

Finite PPP thresholds are bracketed on a $1/16$-dB grid.
Writing their dB values as $s_{\rm P}$ (exact with PPP), $s_{\rm M}$ (exact with mean) and $s_{\rm G}$ (\eqref{eq:gaussian} with mean), define
\begin{equation}
 \Delta_I=s_{\rm M}-s_{\rm P},\quad
 \Delta_G=s_{\rm G}-s_{\rm M},\quad
 \Delta_T=s_{\rm G}-s_{\rm P}.
 \label{eq:errors}
\end{equation}
The total is enclosed directly, because
adding the intervals for $\Delta_I$ and $\Delta_G$ would count the width
of $s_{\rm M}$ twice.

\section{Opposing Errors and False Feasibility}
\subsection{Gain and Interference Errors}
In Fig.~\ref{fig:errors}, at $N=20$, $\Delta_I\in[-3.1250,-2.9375]$
dB and $\Delta_G\in[3.8125,3.9375]$ dB partially compensate, giving
$\Delta_T\in[0.7500,0.9375]$ dB. At $N=24$,
$\Delta_T\in[-0.7500,-0.5625]$ dB. No claim is made at intervening
integers.
With Nakagami-$m$ useful fading in the same environment, the gain
error is smaller at $N=20,24$ for $m\in\{2,3\}$. For $m=2$ the total
error changes sign between the tested sizes $N=8$ ($[2.1875,2.4375]$ dB)
and $N=12$ ($[-1.4375,-1.1875]$ dB); for $m=3$ it is negative at all
eight tested sizes. At the six tested sizes $N\ge16$, the Nakagami-matched \eqref{eq:gaussian}-plus-mean rule underestimates the required SNR by $2.4$--$4.6$ dB for $m=2$ and $3$.

Bouleau's error calculus~\cite{Bouleau2003} approximates the mean-rule outage bias $\Delta$ by a curvature term
\begin{equation}
 \Delta=\E g(J)-g(\E J)
 \simeq\tfrac12 g''(\E J)\operatorname{Var}(J).
 \label{eq:curvature}
\end{equation}
For C-INR, $g(x)=F_W(d_i(1+x)/\rho)$. The certified curvature sign agrees with
that of the bias in all 64 user-1 cases at targets $10^{-4}$ and $10^{-5}$, each at the first certified feasible SNR of the PPP reference;
since all these biases are positive, this is a consistency check rather
than a discriminating test.
However, for the certified feasible evaluation at $3.25$ dB ($N=20$,
$\lambda=0.1$, $\eta=4$, target $10^{-5}$), the curvature term accounts
for only $0.65$--$0.70\%$ of the actual \emph{outage} bias. The enclosures do not
bound the higher-order Taylor terms at the full interference fluctuation.

On Poisson space, the jump generator represents the bias exactly.
Let $Y$ be the driftless subordinator with $Y_1\overset{d}{=}J$; its L\'evy measure
$\nu(dx)=(2\pi\lambda/\eta)x^{-1-2/\eta}\Gamma(1+2/\eta,x)\,dx$ has
infinite activity and finite variation; $\Gamma(a,x)$ is the upper incomplete gamma function. Dynkin's formula on truncations
of $\nu$, then dominated convergence, give exactly
\begin{equation}
 \begin{split}
 \Delta&=\int_0^1\E\,\mathcal A g\bigl(Y_t+(1-t)\E J\bigr)\,dt,\\
 \mathcal A g(y)&=\int[g(y+x)-g(y)-xg'(y)]\,\nu(dx).
 \end{split}
 \label{eq:generator}
\end{equation}
For the C-INR function above, $g\in C_b^6([0,\infty))$
and $\mathcal A$ is the generator of the centered process $Y_t-t\E J$.
Equation \eqref{eq:curvature} uses the second-order local Taylor
approximation $\mathcal D=\frac12\kappa_2\partial^2$ to $\mathcal A$,
with $\kappa_2=\operatorname{Var}J$, and freezes the argument at $\E J$.
The second-order generator approximation
$\Delta^{(2)}=\mathcal A g(\E J)+\frac12\mathcal A^2g(\E J)$ is certified
within a factor two of $\Delta$ in all 64 evaluations, without a
series-remainder bound or a convergence claim. It needs $\nu$-integrals
of $g$, so it explains the bias rather than speeding up its evaluation.

We retain the exact near-interferer distribution without a generator expansion.
Jump--drift--diffusion approximations \cite{BGR2018,AR01}, near/far
interference treatments \cite{CA09,qin} and exact dominant-interferer
treatment with a Gaussian residual approximation and Berry--Esseen
bounds \cite{CD17} motivate our fixed radial split. Krishnan \emph{et al.}\ \cite{krishnan} estimate
coverage by conditional Monte Carlo with Laplace-functional tail
averaging, noting Jensen-type bias in mean-field closures.
Let $J_{[a,b)}$ be the contribution of
interferers with $a\le|x|<b$, and
\begin{equation}
 J_3=J_{[1,r_1)}+\E J_{[r_1,\infty)}
     +\sqrt{\operatorname{Var}J_{[r_2,\infty)}}\;Z',
 \label{eq:hybrid}
\end{equation}
with $Z'\sim\mathcal N(0,1)$ independent of the near interferers and of the
thermal factor. The radii $(r_1,r_2)=(2,4)$ are the smallest powers of
two keeping at least $90\%$ of $\operatorname{Var}J$ in the exact near-interferer term and leaving at most $10^{-4}$ of its third cumulant beyond $r_2$,
for $\eta=3$ and $4$. We fixed them after the preceding analysis, before computing the hybrid. Let
$\Delta_3=\E\tilde g(J_3)-g(\E J)$, where $\tilde g$ extends $g$ with
values in $[0,1]$. We recover the distribution of $J_3$ on $\mathbb R$ by bilateral Laplace inversion \cite{cai}, with analytic bounds on both alias
directions, the negative one through the Gaussian tail; at the tested
$(\lambda,\eta)$ the enclosure includes $\Prb(J_3<0)<1.6\cdot10^{-19}$.

For the same 64 evaluations, bilateral inversion for \eqref{eq:hybrid} gives $\Delta_3/\Delta\in[0.9032,1.0225]$, whereas the curvature term is within a factor of two in only 7. The ratio bounds the outage bias, not
the SNR threshold, and does not follow from the semigroup bounds of
\cite{BGR2018}. For these 64 evaluations, a direct certified bound shows that
removing the Brownian term changes $\Delta_3$ by at most $0.82\%$ of
$\Delta$. Most of the recovered bias therefore comes from the exact near interferers and far mean; the ratio enclosure remains specific to $J_3$.

\subsection{False Feasibility Under C-TX}
For C-TX in the same environment at $N=20$, the exact floor lies in $[1.5887,1.6155]\,10^{-4}$, so $10^{-5}$ is impossible at
\emph{every} SNR. Mean interference nevertheless predicts a finite threshold, $[0.1875,0.2500]$ dB with exact useful
gain and $[6.8750,7.0000]$ dB with \eqref{eq:gaussian}.
The \eqref{eq:gaussian} comparator with Gamma interference also identifies an above-target
floor here, without implying a universal conservative bias.
With Nakagami $m=2$ the false feasibility persists at $N=20$ (exact floor
$[1.2906,1.4341]\,10^{-5}$), also for Gamma interference with exact gain, but not for $m=3$.
For Rayleigh fading, the propagated
Chernoff floor, $[6.4233,6.6764]\,10^{-4}$, cannot certify feasibility.

Let $Q$ be the received power of the geometrically nearest interferer
and $J_{\max}$ the largest received power, so $Q\le J_{\max}\le J$ and $\Prb(J_{\max}\ge x)=1-e^{-\Lambda(x)}$,
$\Lambda(x)=(2\pi\lambda/\eta)\,x^{-2/\eta}\,\Gamma(2/\eta,x)$.
As $F_W$ is increasing, $F_W(d_ix)\Prb(Q\ge x)\le\E F_W(d_iJ_{\max})\le P_{i,\mathrm{floor}}$.
At $N=20$, the first product with $x=4$ exceeds $1.9465\,10^{-5}$, so this elementary bound already proves infeasibility at every SNR. The decisive case is $N=24$ (Table~\ref{tab:floor}): the exact C-TX floor is above the target.
With exact gain, Gamma interference gives a finite threshold in
$[4.50,4.75]$ dB, falsely predicting feasibility. Using \eqref{eq:gaussian} raises the Gamma floor above the target,
masking this false feasibility. The $J_{\max}$ lower bound cannot certify the exclusion.

\begin{table}[!h]
\caption{C-TX, $N=24$, $\lambda=0.1$, $\eta=4$, user 1,
target $10^{-5}$. Values are in units of $10^{-5}$.}
\label{tab:floor}
\centering\footnotesize
\setlength{\tabcolsep}{3pt}
\renewcommand{\arraystretch}{1.13}
\begin{tabular}{@{}lll@{}}
\hline
Model or bound & Certified value & Decision\\
\hline
Exact gain + PPP & $[1.7136,1.7478]$ & excluded\\
Exact gain + Gamma & $[0.5629,0.5757]$ & falsely feasible\\
\eqref{eq:gaussian} + Gamma & $[1.639,1.668]$ & excluded\\
$\E F_W(d_1J_{\max})$ & $<0.9784$ & inconclusive\\
\hline
\end{tabular}
\par\smallskip
\begin{minipage}{\columnwidth}\footnotesize
Only row 1 certifies PPP infeasibility. Row 4 gives an upper bound on a lower bound for the PPP floor.
\end{minipage}
\end{table}

\section{Scope and Numerical Precision}
Of the 384 exact-PPP cases, 240 have attainable targets and 144 are
excluded by their floors. With exact gain and Gamma interference, threshold
errors (Gamma minus PPP) are certified negative in 141 of the 192
C-INR cases and undecided in 51. Under C-TX, Gamma recovers
139 of the 144 exclusions and falsely declares five feasible. The 48 finite
comparisons give 19 negative and 29 undecided signs. With Gamma
interference, \eqref{eq:gaussian} restores four of these five exclusions through
gain-error compensation, but gives one false feasibility and one false
infeasibility over all C-TX cases. Finite dB errors concern targets
attainable under both models.

Under C-INR, the mean rule can overestimate outage. As $\rho\to\infty$, conditioning
on $(A_N,J)$ and dominated convergence, using
$\E A_N^{-1}\le\E(E_1F_1)^{-1/2}=\pi$ and
$\Prb(|10+Z|\le u)\le 2u\,\phi(0)$, give
\begin{equation}
 P_{i,\mathrm{INR}}(\rho)\sim2\phi(10)\,\E[A_N^{-1}]\,
 \E\sqrt{1+J}\,\sqrt{101d_i/\rho},
 \label{eq:ultimate}
\end{equation}
and the same with $\sqrt{1+\E J}$ for the mean rule. Under the adopted
squared-Gaussian harvesting model, whose temperature difference has a
positive density at zero, the ultimate diversity order is thus $1/2$
for every $N$, as for $J=0$, and by strict concavity of the square root the mean rule
overestimates outage at all sufficiently high SNR. The first reversal SNR is not identified. The Zenodo archive holds a certified $140$ dB example.

Across the 240 attainable exact-PPP cases,
the relative width $(P_+-P_-)/P_-$ is at most $4.122\%$.
Without PPP, the 48 thermal Rayleigh evaluations ($N=8,16,32,64$,
both users, $10^{-3}$--$10^{-8}$) give widths below $0.8\%$, a bound specific to these evaluations.
In tests on one 1.8-GHz Xeon Silver 4108 core (160 bits, $N=20,24$),
exact/Chernoff setup took at most $961$ s per size. With banks loaded,
C-INR evaluations took at most $10$ ms and searches at most $122$ ms. These deterministic enclosures hold for the stated model,
inputs and outward-rounding rules. They are not confidence intervals and
do not establish physical validity.

\section{Conclusion}
In the tested regimes, two-sided certification separates opposing power-sizing errors and proves floor exclusions, including one missed by the largest-interferer bound. C-TX designs whose certified floor lower bound exceeds the target should be excluded first. When available, the upper endpoint
of a certified Chernoff-rule bracket gives a sufficient SNR.

\noindent\emph{Code and data:} Zenodo, \href{https://doi.org/10.5281/zenodo.24450534}{doi:10.5281/zenodo.24450534}. Extracted from this archive after submission, the Python package \texttt{certoutage} reproduces Table~II (\href{https://doi.org/10.5281/zenodo.24509256}{doi:10.5281/zenodo.24509256}).


\begin{thebibliography}{20}
\bibitem{boujemaa}
H. Boujemaa, M. Alhussein, and G. Rekaya,
``Reconfigurable intelligent surfaces (RIS) using NOMA with thermal
energy harvesting,'' \emph{Signal, Image Video Process.}, vol.~19,
Art.~no.~419, 2025, doi:~10.1007/s11760-025-03996-x.

\bibitem{atapattu}
S. Atapattu, R. Fan, P. Dharmawansa, G. Wang, J. Evans, and
T. A. Tsiftsis, ``Reconfigurable intelligent surface assisted two-way
communications: Performance analysis and optimization,''
\emph{IEEE Trans. Commun.}, vol.~68, no.~10, pp.~6552--6567,
2020, doi:~10.1109/TCOMM.2020.3008402.

\bibitem{xie}
Z. Xie, W. Yi, X. Wu, Y. Liu, and A. Nallanathan, ``STAR-RIS aided NOMA
in multicell networks: A general analytical framework with Gamma
distributed channel modeling,'' \emph{IEEE Trans. Commun.}, vol.~70,
no.~8, pp.~5629--5644, 2022, doi:~10.1109/TCOMM.2022.3186409.

\bibitem{trigui}
I. Trigui, W. Ajib, W.-P. Zhu, and M. Di Renzo,
``Performance evaluation and diversity analysis of RIS-assisted
communications over generalized fading channels in the presence of
phase noise,'' \emph{IEEE Open J. Commun. Soc.}, vol.~3,
pp.~593--607, 2022, doi:~10.1109/OJCOMS.2022.3160722.

\bibitem{luna}
M. C. Luna Alvarado \emph{et al.}, ``On the exact sum of products
of Rayleigh envelopes with application to RIS,''
\emph{IEEE Wireless Commun. Lett.}, vol.~13, no.~8,
pp.~2145--2149, 2024, doi:~10.1109/LWC.2024.3403391.

\bibitem{wang}
T. Wang, G. Chen, J. P. Coon, and M.-A. Badiu,
``Chernoff bound and saddlepoint approximation for outage probability
in IRS-assisted wireless systems,'' in \emph{Proc. IEEE Globecom
Workshops}, 2021, pp.~1--5, doi:~10.1109/GCWkshps52748.2021.9681955.

\bibitem{bulla}
A. Bulla and S. M. Shah, ``Outage analysis of energy efficiency in
a finite-element-IRS aided communication system,''
\emph{Phys. Commun.}, vol.~54, Art.~no.~101776, 2022,
doi:~10.1016/j.phycom.2022.101776.

\bibitem{anade}
D. Anade, J.-M. Gorce, P. Mary, and S. M. Perlaza,
``An upper bound on the error induced by saddlepoint approximations---
applications to information theory,'' \emph{Entropy}, vol.~22,
no.~6, Art.~no.~690, 2020, doi:~10.3390/e22060690.

\bibitem{garcia}
F. D. Almeida Garc\'ia, F. du Pin Calmon, and
J. C. Silveira Santos Filho, ``Enhanced channel characterization
for RIS-aided systems over Nakagami-$m$ fading channels,''
\emph{IEEE Trans. Veh. Technol.}, vol.~75, no.~4,
pp.~6207--6223, 2026, doi:~10.1109/TVT.2025.3622360.

\bibitem{baccelli}
F. Baccelli and B. B{\l}aszczyszyn, ``Stochastic geometry and
wireless networks: Volume I theory,'' \emph{Found. Trends Netw.},
vol.~3, nos.~3--4, pp.~249--451, 2009, doi:~10.1561/1300000006.

\bibitem{inaltekin}
H. Inaltekin, ``Gaussian approximation for the wireless multi-access
interference distribution,'' \emph{IEEE Trans. Signal Process.},
vol.~60, no.~11, pp.~6114--6120, 2012.

\bibitem{khabou}
M. Khabou and G. L. Torrisi, ``Gaussian approximation and moderate
deviations of Poisson shot noises with application to compound
generalized Hawkes processes,'' \emph{Adv. Appl. Probab.}, vol.~57,
pp.~305--345, 2025, doi:~10.1017/apr.2024.51.

\bibitem{qin}
Y. Qin, M. A. Kishk, and M.-S. Alouini, ``A dominant interferer plus
mean field-based approximation for SINR meta distribution in wireless
networks,'' \emph{IEEE Trans. Commun.}, vol.~71, no.~6,
pp.~3663--3678, 2023.

\bibitem{abate}
J. Abate and W. Whitt, ``The Fourier-series method for inverting
transforms of probability distributions,'' \emph{Queueing Syst.},
vol.~10, nos.~1--2, pp.~5--87, 1992, doi:~10.1007/BF01158520.

\bibitem{arb}
F. Johansson, ``Arb: Efficient arbitrary-precision midpoint-radius
interval arithmetic,'' \emph{IEEE Trans. Comput.}, vol.~66,
no.~8, pp.~1281--1292, 2017, doi:~10.1109/TC.2017.2690633.

\bibitem{mpfi}
N. Revol and F. Rouillier, ``Motivations for an arbitrary precision
interval arithmetic and the MPFI library,'' \emph{Reliable Comput.},
vol.~11, no.~4, pp.~275--290, 2005.

\bibitem{Bouleau2003}
N. Bouleau, \emph{Error Calculus for Finance and Physics:
The Language of Dirichlet Forms}. Berlin, Germany: de Gruyter,
2003, pp.~1--2, 9.

\bibitem{BGR2018}
V. Bally, D. Goreac, and V. Rabiet, ``Regularity and stability for the
semigroup of jump diffusions with state-dependent intensity,''
\emph{Ann. Appl. Probab.}, vol.~28, no.~5, pp.~3028--3074, 2018,
doi:~10.1214/18-AAP1382.

\bibitem{AR01}
S. Asmussen and J. Rosi\'nski, ``Approximations of small jumps of
L\'evy processes with a view towards simulation,''
\emph{J. Appl. Probab.}, vol.~38, no.~2, pp.~482--493, 2001.

\bibitem{CA09}
V. Chandrasekhar and J. G. Andrews, ``Uplink capacity and interference
avoidance for two-tier femtocell networks,'' \emph{IEEE Trans. Wireless
Commun.}, vol.~8, no.~7, pp.~3498--3509, 2009.

\bibitem{CD17}
V. V. Chetlur and H. S. Dhillon, ``Downlink coverage analysis for a
finite 3-D wireless network of unmanned aerial vehicles,''
\emph{IEEE Trans. Commun.}, vol.~65, no.~10, pp.~4543--4558, 2017.

\bibitem{krishnan}
S. R. Krishnan, J. Farooq, K. V. Mishra, X. Liu, U. Pillai, and
T. S. Rappaport, ``Rao--Blackwellized coverage estimation in Poisson
networks: A high-fidelity hybrid framework,'' \emph{IEEE Commun. Lett.},
vol.~30, pp.~3134--3138, 2026, doi:~10.1109/LCOMM.2026.3730490.

\bibitem{cai}
N. Cai, S. G. Kou, and Z. Liu, ``A two-sided Laplace inversion algorithm
with computable error bounds and its applications in financial
engineering,'' \emph{Adv. Appl. Probab.}, vol.~46, no.~3, pp.~766--789,
2014, doi:~10.1239/aap/1409319559.
\end{thebibliography}
\end{document}